# How To Identify Plasmons from the Optical Response of Nanostructures

Runmin Zhang,[†] Luca Bursi,[‡,§,◇] Joel D. Cox,[∥] Yao Cui,[†,⊥] Caroline M. Krauter,[#] Alessandro Alabastri,[†] Alejandro Manjavacas,[▽] Arrigo Calzolari,[§] Stefano Corni,[*,§,○] Elisa Molinari,[‡,§] Emily A. Carter,[◆] F. Javier García de Abajo,[*,∥,¶] Hui Zhang,[*,†] and Peter Nordlander[*,†]

[†]Laboratory for Nanophotonics and the Department of Physics and Astronomy, MS61 and [⊥]Department of Chemistry, Rice University, Houston, Texas 77005, United States
[‡]Dipartimento di Fisica, Informatica e Matematica-FIM, Università di Modena e Reggio Emilia, I-41125 Modena, Italy
[§]Istituto Nanoscienze, Consiglio Nazionale delle Ricerche CNR-NANO, I-41125 Modena, Italy
[∥]ICFO-Institut de Ciencies Fotoniques, The Barcelona Institute of Science and Technology, 08860 Castelldefels, Barcelona, Spain
[#]Department of Mechanical and Aerospace Engineering and [◆]School of Engineering and Applied Science, Princeton University, Princeton, New Jersey 08544-5263, United States
[▽]Department of Physics and Astronomy, University of New Mexico, Albuquerque, New Mexico 87131, United States
[○]Dipartimento di Scienze Chimiche, Università di Padova, I-35131 Padova, Italy
[¶]ICREA-Institució Catalana de Reserca i Estudis Avançats, Passeig Lluís Companys 23, 08010 Barcelona, Spain

**S** Supporting Information

**ABSTRACT:** A promising trend in plasmonics involves shrinking the size of plasmon-supporting structures down to a few nanometers, thus enabling control over light−matter interaction at extreme-subwavelength scales. In this limit, quantum mechanical effects, such as nonlocal screening and size quantization, strongly affect the plasmonic response, rendering it substantially different from classical predictions. For very small clusters and molecules, collective plasmonic modes are hard to distinguish from other excitations such as single-electron transitions. Using rigorous quantum mechanical computational techniques for a wide variety of physical systems, we describe how an optical resonance of a nanostructure can be classified as either plasmonic or nonplasmonic. More precisely, we define a universal metric for such classification, the generalized plasmonicity index (GPI), which can be straightforwardly implemented in any computational electronic-structure method or classical electromagnetic approach to discriminate plasmons from single-particle excitations and photonic modes. Using the GPI, we investigate the plasmonicity of optical resonances in a wide range of systems including: the emergence of plasmonic behavior in small jellium spheres as the size and the number of electrons increase; atomic-scale metallic clusters as a function of the number of atoms; and nanostructured graphene as a function of size and doping down to the molecular plasmons in polycyclic aromatic hydrocarbons. Our study provides a rigorous foundation for the further development of ultrasmall nanostructures based on molecular plasmonics.

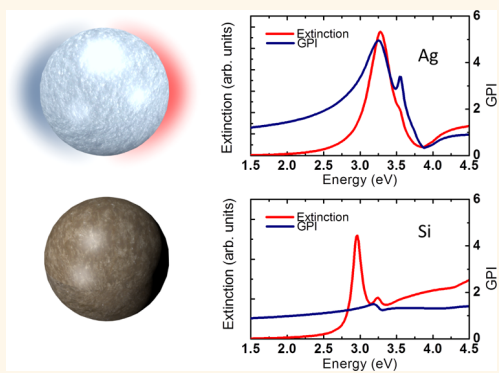

**KEYWORDS:** *plasmon, collective excitation, RPA, TDDFT, Mie theory, jellium model*

**P**lasmons, the collective electron oscillations in metallic nanostructures,[1] play a major role in a variety of applications due to their ability to confine light down to subwavelength volumes. These applications include chemical and biological sensing,[2−5] waveguiding,[6] energy-transfer processes,[7,8] light harvesting,[9,10] photodetection,[11] photocatalysis,[11,12] and photothermal cancer therapy.[13] Plasmons in noble metal structures are well described by classical electromagnetism when their size exceeds a few nanometers. However, quantum mechanical and finite-confinement effects emerge in morphologies including nanometer-sized gaps, tips, and edges.[14−19] These observations have created the subfield of quantum plasmonics, where quantum mechanical effects can be optically probed and exploited for active control of plasmonic resonances.[20,21] In this context, a great deal of work has been devoted recently to study the plasmonic properties of ultrasmall nanostructures containing <1000 conduction electrons,[14,22−25] because such structures provide deep subwavelength confinement and present a large surface-to-volume ratio, an important parameter in sensing and photocatalysis.[23,26−28] In this paper, we are concerned primarily with localized surface plasmon resonances (LSPRs) that appear in finite nanostructures and, in contrast to surface plasmon polaritons (SPPs) typical of extended structures, are characterized by a discrete excitation spectrum. However, our conclusions are most likely valid for both types of plasmonic modes.

Identification of plasmons is challenging in small systems (*e.g.*, clusters or molecules), particularly when only a few electrons are involved[29−36] as, for example, in molecules consisting of only a few tens of atoms.[37−39] Experimentally, optical resonances have been studied in graphene-like polycyclic aromatic hydrocarbon (PAH) molecules,[37,40] which present several characteristic properties of LSPRs. However, arguments regarding their classification as plasmons still persist. When the number of charge carriers in a system decreases, the energy gap between the quantized electronic states increases, resulting in blurred boundaries between conventional plasmonic behavior and electron−hole (e−h) pair transitions, which also appear as optical resonances.[31] Within a quantum mechanical picture, plasmons are usually described as coherent superpositions of certain e−h pair quantum states (Slater determinants). In contrast, pure single-electron transitions do not exhibit such coherent behavior.

Several theoretical studies on few-electron nanoparticles (NPs) or molecular-scale systems have provided important insights into the origin and emergence of plasmons. For instance, Gao and co-workers studied linear atomic chains using time-dependent density functional theory (TDDFT)[34,35] and showed how collective excitations emerge as the length of the chain increases. Bryant and co-workers compared TDDFT results with exact calculations for small model linear systems.[41,42] More recently, Fitzgerald *et al.* explored single and coupled sodium atomic chains as ultrasmall molecular plasmonic nanoantennas.[36] Bernadotte *et al.* and Krauter *et al.* proposed a scaling procedure of the interaction between electrons in various first-principles calculation frameworks[30,32] in order to show that plasmons and single-electron transitions evolve in different ways when varying the strength of such interaction. Jain studied the physical nature of few-carrier plasmon resonances using a model in which the electrons were confined to a potential box.[33] The random-phase approximation (RPA) was adopted to investigate doped nanocrystals[26] as well as graphene nanoislands[43] with increasing carrier density, showing both the evolution of plasmon modes and quantum finite-size effects. These studies focused on the microscopic nature of plasmons and relied on elaborate quantum mechanical descriptions to characterize the plasmonic nature of optical excitations, which, due to their computational cost, cannot be extended to larger systems.

Clearly, a more universal approach for classifying plasmonic behavior and distinguishing such excitations from single-electron transitions would be highly desirable. Such an approach should ideally be feasible for systems of arbitrary composition and number of charge carriers. Very recently, Bursi *et al.* proposed a "plasmonicity index" (PI) to characterize and quantify plasmonic behavior.[44] This pioneering method was based on the Coulomb potential induced in the NP upon external illumination and paved the way toward a universal semiclassical metric for identifying plasmonic behavior following a simple, specific procedure that does not depend on the details of the quantum mechanical model used for the analysis. In this context, we also note that Yan and Mortensen have introduced a nonclassical-impact parameter within the RPA in order to characterize the degree of nonclassical effect in plasmon resonance dynamics.[45] However, these metrics cannot be applied to classical systems, and because the PI was not dimensionless, it did not enable a direct comparison of the "plasmonicity" for optical excitations in different structures.

In this paper, we first adopt the RPA[26,46−48] framework to discuss the fundamental difference between plasmon resonances and other types of optical excitations. We study the evolution of plasmon resonances and single-electron transitions in absorption spectra as a function of size and number of conduction electrons for a wide variety of systems, ranging from small metallic nanospheres and metal clusters described within the jellium model to graphene nanostructures and polycyclic aromatic hydrocarbons in which a tight-binding model based upon localized orbitals provide a better description of valence electrons. We introduce an improved metric, the generalized plasmonicity index (GPI) to distinguish plasmons from single-electron transitions and show its general validity for discriminating plasmons from other types of optical modes in systems of varying size, number of electrons, and chemical nature. Unlike the PI metric,[44] the GPI is dimensionless and can be readily used both with any quantum chemistry approach and for systems that are well-described by classical electromagnetism. The GPI not only provides a simple means of answering the question, how many electrons are required to support a plasmonic collective mode in a given class of systems, but also offers an intuitive perspective to understand the fundamental properties of plasmon modes.

The organization of our paper is as follows: First we discuss the qualitative difference between plasmons and single-particle excitations, and we show how they can be distinguished within the RPA. What follows then is the core of the paper: We introduce the GPI, which provides a universal quantitative metric for the plasmonic character of optical excitations. We show hence how the GPI can be calculated at several different levels of theory (classical electrodynamics, jellium, first-principles) and then use it to study the emergence of plasmonic behavior as the number of electrons or atoms in a structure increases. We conclude with an investigation of the plasmonicity of optical excitations in nanostructured graphene and in polycyclic aromatic hydrocarbons for different levels of doping.

## THEORY

**Plasmons and Single-Particle Excitations.** A metallic NP can be described simplistically as a conduction electron gas confined to a lattice of positive ions. In this picture, LSPRs emerge as collective excitations involving incompressible deformations of the conduction electron gas with respect to its uniform equilibrium distribution.[49] Figure 1 conceptually illustrates the induced charges associated with a dipolar plasmon in a spherical NP. The displacement of the conduction electrons exposes the positive background on one side of the NP and results in negative surface charges on the other side. Consequently, Coulomb interactions within and between the induced charge distribution create a restoring force that, combined with the kinetic energy of the conduction electron gas, which effectively acts like a repulsive force, defines a harmonic oscillator. The coherent motion of many conduction electrons in a NP contributes to a large dipole moment and accordingly is responsible for the strong coupling of LSPRs to light. The significant amount of surface charges that can be induced from resonant excitations of a LSPR mode is responsible for the large electric-field amplitude enhancements induced outside the NP surfaces, which can easily exceed factors of several hundreds and cause inelastic optical effects such as surface-enhanced Raman scattering (SERS) from a single molecule adsorbed on the NP surface.[4]

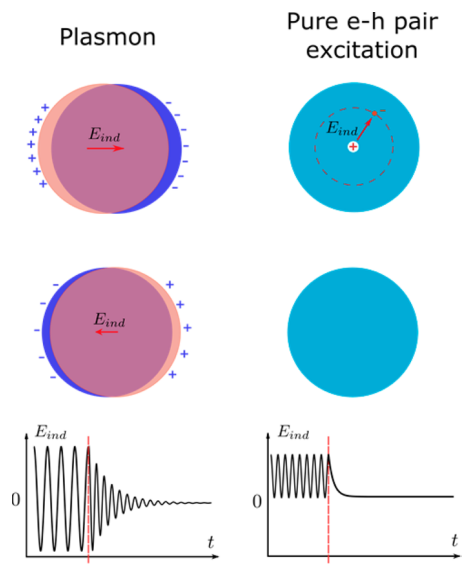

Figure 1. Schematic illustration of the differences between a collective excitation such as a dipolar plasmon (left) and a pure e–h pair excitation (right). The plasmon involves the coherent motion of all conduction electrons in the NP and can be roughly modeled as rigid displacements of the conduction electron gas. This motion induces strong surface charges at the interface, which in turn induce a significant field $E_{ind}$ (the depolarization field) inside the NP. Therefore, plasmonic excitations continue as damped oscillations after the driving external field is turned off (vertical dashed line in the lower panels). In contrast, an e–h pair has no significant induced field beyond the screened internal Coulomb interaction between the hole and the electron: Rabi oscillations induce a time-dependent dipolar field as the external field persists, while after decay the electron recombines with the hole without ringing (schematically illustrated by an exponential attenuation instead of a step function).

Just as the plasmon-induced surface charges generate large field enhancements outside the supporting NPs, they are also responsible for an induced field $E_{ind}$ (the depolarization field) across the interior. In contrast, in pure single-electron transitions, a charge carrier is excited to an unoccupied state, and the induced internal field is weaker. Depending on the strength of the interaction between the electron and the corresponding hole, the excitation can be long-lived and is often referred to as an exciton. An exciton also includes correlated motion between the electron and its corresponding hole together with other electrons involved in the screened e–h pair; however the phenomenon is not collective in the same sense as in a plasmon:[48] A plasmonic oscillation strongly depends on the electronic motion spreading across the NP on which it is sustained, which is why this collective phenomenon is strongly influenced by the NP geometry; in contrast, a single-electron transition is a localized event that does not strongly depend on other states created elsewhere in the NP. While in principle all the electrons participate in screening the e–h pair, low-density excitons interact weakly and can be considered to be independent quasi-particles coupled mainly through Coulomb interactions constrained to the e–h region, as depicted in Figure 1.

Figure 1 also illustrates the fundamental difference in temporal behavior between plasmons and pure e–h pair excitations in a semiclassical approximation. During irradiation with light that is slightly red-shifted (blue-shifted) from the plasmon resonance frequency $\omega_p$, the electron gas oscillates with its dipole in phase (out of phase) with the applied electric field. When the illumination ceases, the plasmon oscillations continue, aided by Coulomb interactions, but with decreasing amplitudes due to damping (lifetime $\tau$). The latter is due to the progressive dephasing of the oscillations of the e–h pairs that collectively form the plasmon, which finally decay incoherently by e–h recombination. The number of self-sustained oscillations of the electron cloud after the incident light is turned off is proportional to the quality factor $Q = \omega_p \tau$ of the plasmon resonance (the number of oscillations is a factor of $2\pi$ smaller than $Q$ when the induced-field intensity has decreased by a factor of $e$).[50] In contrast, for a pure e–h pair excitation, although the internal field is modified by the dipolar field generated by Rabi oscillations while the external illumination persists, there is no dephasing step after the illumination ceases because a single e–h pair is present, and the electron recombines with the hole in a monotonic manner, assuming the electron and hole are still coupled and have not separated. In conclusion, the motion of the e–h pairs in the collective plasmon mode is to a large extent determined by its self-induced surface charges, while the pure e–h pair dynamics of an exciton are controlled mainly by the external field.

**Identification of Plasmons within the Random-Phase Approximation.** Perhaps the most transparent way of describing the difference between plasmons and single-electron transitions is through the RPA. For simplicity we illustrate this here in momentum space $\mathbf{k}$ (i.e., using a plane-wave decomposition), assuming a homogeneous system (i.e., translational invariance). Although this basis is not optimal for numerical calculations of finite nanostructures, the mathematical description simplifies considerably compared to a real-space description. Optical absorption in a nanostructure is determined by the motion of the induced charges with respect to the incident field and can be calculated from the imaginary part of the temporal Fourier transform of the induced charge distribution (i.e., working in frequency space $\omega$). Within linear response theory,[51] the induced charge distribution $\delta n(\mathbf{k}, \omega)$ can be written as

$$\delta n(\mathbf{k}, \omega) = \chi^0(\mathbf{k}, \omega) v_{tot}(\mathbf{k}, \omega) = \chi^0(\mathbf{k}, \omega)[v_{ext}(\mathbf{k}, \omega) + v_{ind}(\mathbf{k}, \omega)] \quad (1)$$

where $\chi^0$ is the so-called non-interacting (or independent-electron) susceptibility and $v_{ext}$ is the applied external potential, which for an incident electromagnetic field is the associated quasistatic electric potential, whereas $v_{ind}$ is the induced potential, responsible for both the external field enhancement in the vicinity of the NP and the internal depolarization field. The use of the quasistatic approximation is justified by the small size of the plasmonic nanostructures as compared to their resonant wavelengths. The susceptibility $\chi^0$ can be calculated using perturbation theory directly from the electronic wave functions of the system,[51] and expressed as a sum of Lorentzian terms, one per e–h pair excitation, weighted by the transition matrix elements connecting electron and hole wave functions via the external exciting field. For a homogeneous electron gas, which is a good model to describe simple metals such as sodium and aluminum, $\chi^0$ can be expressed in an explicit closed-form expression that is often referred to as the Lindhard susceptibility.[52] For light excitation, the wave vector and frequency are related through the relation $\omega = ck$. Now, for single-particle transitions, the induced potential $v_{ind}$ is typically small and can be neglected, so that single-particle resonances are directly revealed by the poles of $\chi^0$. In practice, these poles occur at complex frequencies, with negative imaginary parts that reflect their damping. For this reason, $\chi^0$ does not diverge at real frequencies, but it can become very large near one such resonant mode.

In contrast to pure e–h pair excitations, in a plasmonic system near resonance, the induced potential $v_{ind}$ can be large compared with the external potential, thus contributing significantly to $\delta n$. In the RPA, one assumes the induced electric field to be directly related to the induced charge density via Gauss's law:

$$v_{ind}(\mathbf{k}, \omega) = \frac{4\pi e^2}{k^2} \delta n(\mathbf{k}, \omega) \quad (2)$$

With this expression inserted on the right-hand side of eq 1, the induced charge density can be written as

$$\delta n(\mathbf{k}, \omega) = \frac{\chi^0(\mathbf{k}, \omega)}{1 - \frac{4\pi e^2}{k^2}\chi^0(\mathbf{k}, \omega)} v_{ext}(\mathbf{k}, \omega) = \chi(\mathbf{k}, \omega) v_{ext}(\mathbf{k}, \omega) \quad (3)$$

which implicitly defines $\chi$ as the (interacting) RPA susceptibility. In the above analysis of the RPA susceptibility, we neglect electron spin, exchange, and correlation effects. The single-particle poles that were present in $\chi^0$ are not present in $\chi$, which instead has poles when the denominator $1 - \frac{4\pi e^2}{k^2}\chi^0(\mathbf{k}, \omega)$ becomes zero. Near these zeros, $\chi$ typically exhibits fast variations that define new dressed excitations. Depending on the strength of the Coulomb interaction ($\frac{4\pi e^2}{k^2}$ in eq 2), the energies of these transitions can be close to the single-particle transitions (then we have excitonic modes that are only weakly affected by the Coulomb interaction) or significantly different, in which case the new poles correspond to plasmon resonances and are directly enabled by the induced potential $v_{ind}$.

For simple systems in which $\chi$ and $\chi^0$ become analytical (e.g., the non-interacting, homogeneous electron gas), the poles of these functions provide an ideal way of distinguishing plasmons from other types of optical excitations. However, most systems require complicated numerical solutions. For example, in the widely used TDDFT approach applied to atoms and molecules, it is far from trivial to identify among the poles of $\chi$ those that are reminiscent of the poles of $\chi^0$. As a result, one cannot trivially distinguish single-electron transitions from plasmons. A practical way of making this distinction consists of introducing a scaling parameter $\lambda$ in the Coulomb potential $\lambda\frac{4\pi e^2}{k^2}$, leading to the effective susceptibility:

$$\chi^{eff}(\lambda, \mathbf{k}, \omega) = \frac{\chi^0(\mathbf{k}, \omega)}{1 - \lambda\frac{4\pi e^2}{k^2}\chi^0(\mathbf{k}, \omega)} \quad (4)$$

which for $\lambda \to 0$ becomes equal to $\chi^0$ and for $\lambda \to 1$ becomes equal to $\chi$. In this transformation, only the induced Coulomb potential $v_{ind}$ in eq 2 should be scaled, not the Coulomb potential which determines the electronic structure of the system. Such scaling of the induced Coulomb potential can be implemented in TDDFT to switch between $\chi$ and $\chi_0$,[30] which thus provides a means for distinguishing between single-electron transitions and plasmons; a plasmon mode will be a resonance that depends on $\lambda$, while single-electron transitions will be relatively independent of $\lambda$.

**Beyond Qualitative Identification: The Generalized Plasmonicity Index.** Recently, Bursi et al.[44] proposed a plasmonicity index (PI) to quantify the difference between plasmons and single-electron transitions in finite structures. The PI is proportional to the integral of the squared modulus of the induced potential $|v_{ind}(r,\omega)|^2$ over the volume of the NP and thus captures the fundamental difference between plasmons and single-electron transitions discussed in conjunction with eq 3. By plotting the PI as a function of excitation energy, plasmon resonances can thus be identified directly as peaks in a simple and straightforward manner. However, a shortcoming of the PI is that its values are not dimensionless and depend on the size of the system, which complicates the comparison between structures of different size. Also, the PI cannot be used to identify plasmonic behavior in classical systems because the induced charges are then 2D surface charges, resulting in an energy-independent PI with a value determined only by the geometry and size of the NP (see section S1 in the Supporting Information (SI) for more details). To construct a more universal metric for classifying plasmonic behavior, we propose a generalized plasmonicity index (GPI) denoted as $\eta$, which is based on the electrostatic Coulomb energy in the system associated with the induced charges:

$$\eta = \frac{|\int d\mathbf{r} \delta n(\mathbf{r}, \omega) v_{ind}^*(\mathbf{r}, \omega)|}{|\int d\mathbf{r} \delta n(\mathbf{r}, \omega) v_{ext}^*(\mathbf{r}, \omega)|} \quad (5)$$

The superscript "*" denotes the complex conjugate of the corresponding quantity. In the above expression, the normalization (i.e., the denominator) is proportional to the Coulomb energy associated with the interaction between the induced charge and the incident electromagnetic field. In the quasistatic limit, this normalization factor is proportional to the total induced dipole moment $D$ of the system:

$$|\int d\mathbf{r} \delta n(\mathbf{r}, \omega) v_{ext}^*(\mathbf{r}, \omega)| \simeq E_0 D, \quad D = |\int d\mathbf{r} \mathbf{d}(\mathbf{r}, \omega)| \quad (6)$$

where $\mathbf{d}(\mathbf{r},\omega) = \mathbf{r}\delta n(\mathbf{r},\omega)$ is the local induced dipole moment at position $\mathbf{r}$, and $E_0$ is the amplitude of the incident field.

The generalized plasmonicity index $\eta$ is a quantitative measure of the Coulomb energy associated with the oscillating polarization of the system and displays a peak at the plasmon energy because these excitations produce a large induced potential (i.e., $v_{ind} \gg v_{ext}$). In contrast, for a single-electron transition, where $v_{ind}$ has a similar order of magnitude as $v_{ext}$, we would expect a GPI close to 1, but the GPI could in principle also approach 0 if $v_{ind}$ becomes very small due to the absence of polarization mechanisms (more details are given in section S2 in the SI). Remarkably, in the quasistatic classical limit the GPI

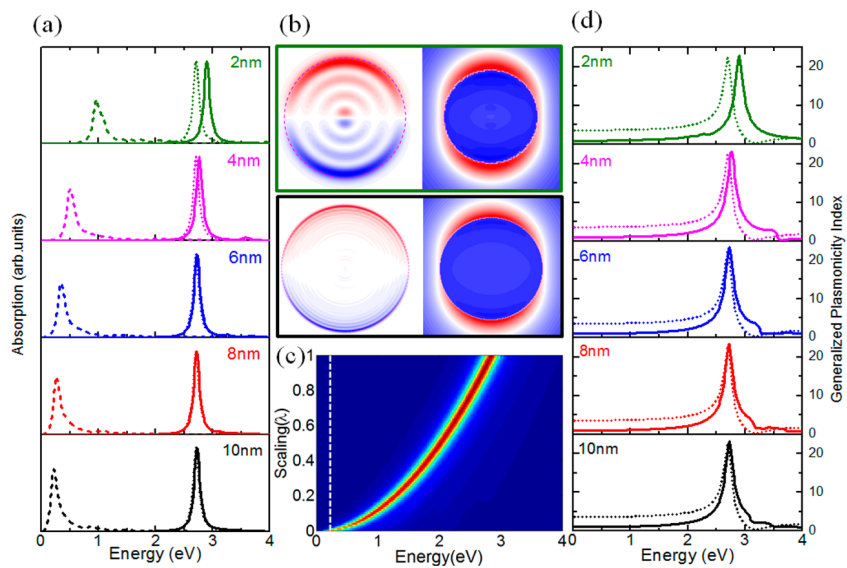

Figure 2. Plasmonicity in small metallic NPs: size dependence. (a) Size evolution of the absorption spectra for gold spheres with diameters from 2 to 10 nm, calculated from the full RPA ($\sigma_{\text{RPA}}(\omega)$, solid curves) and the non-interacting RPA ($\sigma_0(\omega)$, dashed curves) for jellium (5.9 × $10^{22}$ cm$^{-3}$ electron density). Classical Drude−Mie theory is also included for comparison (dotted curves). (b) Induced charge density (left column) and field enhancement (right column) at the plasmon resonance for particles of diameters $D = 2$ nm (top) and $D = 10$ nm (bottom) spheres. (c) Color contour plot of $\sigma_\lambda(\omega)$ for a $D = 8$ nm jellium sphere. The vertical axis represents $\lambda$ varying from 0 to 1 (see main text). The curved streak corresponds to $\sigma_\lambda(\omega)$, while the vertical line indicates the maximum of $\sigma_0(\omega)$. (d) GPI spectra as a function of incident energy for the same spheres as in (a) calculated using the full RPA (solid curves) and classical Drude−Mie theory (dotted curves, multiplied by a factor of 3.6).

relates directly to experimentally accessible quantities (see section S5 in the SI).

**GPI and *ab Initio* Excitation Energies.** The expression of the GPI in eq 5 can be related to a microscopic energy decomposition of the excitation energy at the RPA level that provides further insights into its physical meaning. Starting from the RPA pseudo-eigenvalue equations in matrix form, we can obtain the following expression for the $I^{\text{th}}$ excitation energy, as detailed in Section S3 in the SI:

$$\hbar\omega_I = \sum_{ai}(\epsilon_a - \epsilon_i)(\rho_{ai}^{I*}\rho_{ai}^I + \rho_{ia}^{I*}\rho_{ia}^I)$$
$$+ \iint d\mathbf{r}_1 d\mathbf{r}_2 \frac{\rho_I(\mathbf{r}_1)\rho_I(\mathbf{r}_2)}{|\mathbf{r}_1 - \mathbf{r}_2|} \quad (7)$$

where

$$\rho_I(\mathbf{r}) = \sum_{bj}[\rho_{bj}^I \phi_b(\mathbf{r})\phi_j^*(\mathbf{r}) + \rho_{jb}^I \phi_j(\mathbf{r})\phi_b^*(\mathbf{r})] \quad (8)$$

is the transition density for the excitation $I$, the indices $i,j$ and $a,b$ run over occupied and empty (spin)-orbitals $\phi(r)$ with single-particle energies $\epsilon_i, \epsilon_j$ and $\epsilon_a, \epsilon_b$, respectively, and $\rho_{bj}^I, \rho_{jb}^I$ are the corresponding expansion coefficients of the transition density $\rho_I(\mathbf{r})$.

The first term on the rhs of eq 7 is the single-particle contribution to the excitation energy, being an average of single-electron transition energies ($\epsilon_a - \epsilon_i$) weighted by the transition density matrix expansion elements $\rho_{ai}^I$ and $\rho_{ia}^I$. The second term in the rhs of eq 7 has three features that qualify it as a measure of the plasmonic character of the excitation: (i) it is the only term found in the RPA besides the single-particle contribution, and the RPA is known to describe only single-electron transitions and plasmons; (ii) it is the Coulomb energy associated with the transition charge density and thus is associated with the induced potential that is enhanced in plasmons; (iii) it shifts excitation energies *upward* with respect to the single-particle results, as expected for plasmons.[53,54] Thus, we label this term the *plasmonic energy* of transition $I$:

$$E_{\text{plas},I} = \iint d\mathbf{r}_1 d\mathbf{r}_2 \frac{\rho_I(\mathbf{r}_1)\rho_I(\mathbf{r}_2)}{|\mathbf{r}_1 - \mathbf{r}_2|} \quad (9)$$

This was also the term scaled by $\lambda$ in the approach by Bernadotte *et al.* and Krauter *et al.*[30,32] to identify plasmonic behavior. This scaling approach will be exploited for jellium particles in the Results and Discussion section below. Superficially, the quantities $E_{\text{plas},I}$ and the GPI appear unrelated (GPI is a frequency-dependent quantity based on a response formalism, whereas $E_{\text{plas},I}$ is a single number per each excitation $I$), but they convey the same physics, namely the Coulomb interaction. Indeed, by comparing the GPI evaluated at resonance, $\eta(\omega_I)$, as well as $E_{\text{plas},I}$, it can be shown that they are proportional:

$$\eta(\omega_I) = \frac{E_{\text{plas},I}}{\Gamma} \quad (10)$$

where $\Gamma$ is the damping energy of the transition (see section S4 in the SI for a proof and for a connection between the GPI and the poles of the response function). Eq 10 reveals the microscopic nature of the GPI: a direct measure of the plasmonic contribution to the excitation energy, weighted by the damping of the excitation. The combination of a high Coulomb energy along with a small damping of charge oscillation provides the most pronounced plasmonic transitions, whereas damping of the excitation (i.e., increase of $\Gamma$) degrades the GPI. Eq 10 has the typical form of a Q-factor, representing the Coulomb energy stored in the plasmon divided by the width of the resonance. This quantity can be calculated for dark modes as well.

The connection established in eq 10 allows for a direct calculation of the GPI from any first-principles approach that provides transition densities (instead of charge susceptibilities as obtained from linear-response TDDFT) such as the Casida formulation of TDDFT or wave function-based methods.[55]

## RESULTS AND DISCUSSION

**Implementation of the GPI.** In this subsection, we analyze a wide range of systems and discuss the implementation of the GPI in various computational approaches ranging from TDDFT to classical Mie theory.

**Plasmons in Jellium Nanospheres.** Here, we calculate absorption spectra for jellium nanospheres with parameters (electron density, work function, and lattice background polarizability) chosen to represent Au. We focus on the evolution of the plasmon resonance and the single-electron transitions without considering interband transitions (IBTs) that would be present in real materials. The spectra (see Methods section) calculated using the full RPA with $\chi$ as in eq 3 are denoted $\sigma_{RPA}(\omega)$, while $\sigma_0(\omega)$ refers to single-electron transition spectra calculated using $\chi^0$, and $\sigma_\lambda(\omega)$ identifies the spectra calculated using a real-space version of the scaled $\chi^{eff}(\lambda)$ defined in eq 4. Figure 2a shows the calculated absorption spectra $\sigma_{RPA}$ and $\sigma_0$ for gold jellium nanospheres of diameters ranging from 2 nm (247 electrons) to 10 nm (30,892 electrons) with a damping of 0.12 eV. For comparison, we also show the results from classical calculations using Mie theory for a Drude-model dielectric function with parameters chosen to fit the Johnson and Christy dielectric data[56] for Au in the visible range (neglecting IBTs): $\varepsilon(\omega) = \varepsilon_\infty - \frac{\omega_B^2}{\omega(\omega + i\gamma)}$, with $\varepsilon_\infty = 9.1$, $\omega_B = 9.07$ eV, and a damping $\gamma = 0.12$ eV. The spectra clearly reveal the difference between single-electron and collective plasmon excitations. The $\sigma_{RPA}(\omega)$ spectra have a pronounced plasmon resonance around 2.7 eV, in excellent agreement with the Mie result. The single-electron spectra exhibit an asymmetric absorption band at energies mostly below 1 eV. Both the plasmon and the single-electron transitions undergo a redshift with increasing sphere diameter but appear to saturate beyond 8 nm. These shifts are due to quantum size confinement and agree qualitatively with previous studies of size-dependent plasmon resonances of NPs.[14] To further understand the redshift of the plasmon resonances with increasing size, Figure 2b shows the plasmon-induced charge density and field enhancement at the resonance frequency for diameters of 2 and 10 nm. For the smallest diameter $D = 2$ nm, the induced charges are mostly located inside the NP and distributed in concentric shells, reflecting the discrete electronic structure of the particle. This is in sharp contrast with the classical prediction of a pure surface charge. When the NP size increases, the electronic structure becomes denser, and the induced charge density becomes more prominent at the surface, as expected classically. Interestingly, the plasmon-induced near-field is well developed and agrees with classical theory (not shown) already for the smallest $D = 2$ nm particle. Figure 2c shows the scaled $\sigma_\lambda(\omega)$ spectra for a $D = 8$ nm sphere. The plasmonic feature can also clearly be identified here as the dispersive mode $\omega_\lambda$ that appears at 0.3 eV for $\lambda = 0$ and reaches 2.7 eV for $\lambda = 1$. The other single-electron transitions are essentially independent of $\lambda$. The clear differences between $\sigma_{RPA}(\omega)$ and $\sigma_0(\omega)$ spectra for the nanospheres in Figure 2 make it straightforward to distinguish between single-electron and collective plasmon resonances for the case of jellium nanospheres.

In Figure 2d, we show the calculated GPI spectra (eq 5) for the same jellium spheres as in Figure 2a. As expected, prominent peaks appear in the GPI spectra at energies corresponding to plasmon excitations. In the low-energy region, where single-electron transitions are present, the GPI remains flat with values around unity. Figure 2d also shows the GPI for a sphere described using the classical Drude–Mie model. Again, the plasmon modes can be clearly discerned as distinct peaks. Because the distributions of surface charges are different in RPA and classical Drude–Mie calculations ($\delta$-function distributions in classical theory versus finite distributions of surface charge in the RPA model), the maximum GPI values in the classical results are different. In particular, in the quasistatic classical regime, the GPI reduces to $\eta(\omega) = \left|\frac{\varepsilon(\omega) - 1}{\varepsilon(\omega) + 2}\right|$ (see Section S5 in the SI), where $\varepsilon(\omega)$ is the permittivity of the spherical NP. We observe that the classical result for the GPI is very similar to the RPA result with a pronounced peak at the plasmon energy.

We remark that the GPI is dimensionless and its amplitude at the plasmon resonances for $D = 2-10$ nm is large, thus signaling plasmonic excitations. We conclude that the GPI can identify plasmonic behavior even when strong quantum finite-size effects are present, as in the $D = 2$ nm NP. Because the GPI is expressed in terms of induced charges and Coulomb potentials, which are standard physical quantities, it is a straightforward quantity to calculate in any computational method, whether *ab initio* (density or wave function based), empirical, or based on classical electromagnetism. Importantly, the calculation of the GPI is a postprocessing procedure that can be performed after the time-consuming self-consistent calculations of the response functions are carried out, and it does not require separate calculations of the self-consistent $\sigma_{RPA}(\omega)$ and $\sigma_0(\omega)$ or the calculation of $\sigma_\lambda(\omega)$ spectra for different $\lambda$.

**Probing Classical Mie Modes with the GPI.** We show next that the GPI can also capture the difference between plasmonic and photonic modes in NPs. Photonic modes in large dielectric NPs are currently the subject of considerable research due to their small losses.[57,58] In Figure 3 we compare optical spectra and GPI for Ag and Si nanospheres with diameters of 80 nm. Such large particles are not in the quasistatic regime, and thus, we calculate the GPI using Mie theory to obtain the induced density, which is then directly plugged into the definition of the GPI (eq 5), therefore extending its range of validity to larger systems affected by phase retardation (*i.e.*, different parts of the NP experiencing different fields because of the finite speed of light). The Ag NP presents a dipolar resonance at ~3.3 eV and a quadrupolar mode at ~3.5 eV (Figure 3a). The Si sphere exhibits two Mie resonances: a magnetic dipole around 3.0 eV and an electric dipole around 3.3 eV (Figure 3b). Figure 3c displays the corresponding GPI as a function of incident energy for the two particles. It is reassuring to observe that the GPI spectrum for the metallic Ag sphere has a clear peak and large GPI values ($\eta \gg 1$), clearly showing that the Ag resonances are plasmonic. On the other hand, the GPI for the Si particle does not show a distinct resonance and lingers around unity, which clearly points to its nonplasmonic origin.

Because the GPI measures the plasmonicity of a mode, this metric may be used to quantify how "plasmonic" conventional classical plasmons are. Perhaps even more importantly, it is of interest to pose the question of which classical plasmonic properties contribute to a large GPI. The GPI may then be used as a metric also for classical plasmonic systems. Figure 4 compares the Mie theory calculated extinction cross sections, GPI, and maximum electric-field enhancements for spheres made of Ag, Au, and Al as a function of energy. The solid curves in Figure 4a show the extinction spectra of 50 nm diameter Au and Ag spheres and of 25 and 50 nm diameter Al spheres. The plasmon resonance of the Au sphere overlaps the IBTs and is not particularly intense. This is reflected in the corresponding GPI spectra in Figure 4b, where the GPI value of Au (solid red curves) is much smaller than the GPI value of Ag (solid black curves), because of the damping due to IBTs. The dashed-red curves show the red-shifted spectra for an Au sphere immersed in a dielectric medium. As the plasmon resonance shifts away from the IBTs, it becomes much more pronounced, with a GPI spectrum that clearly signals a plasmonic behavior ($\eta \gg 1$). The spectra for the 50 nm Al sphere also exhibit excellent plasmonic character. Because of the large plasmon frequency of Al, retardation effects are large, and the dipolar resonance is highly damped by radiative losses, thus showing a relatively modest GPI. However, the quadrupolar and octupolar Al resonances, with their significantly smaller radiative damping, exhibit excellent plasmonic behavior with $\eta > 5$. The dashed-blue curves show the spectra for a smaller Al sphere (25 nm diameter). Here the radiative damping is significantly reduced, thus resulting in significantly better plasmonic behavior also for the Al dipolar resonance.

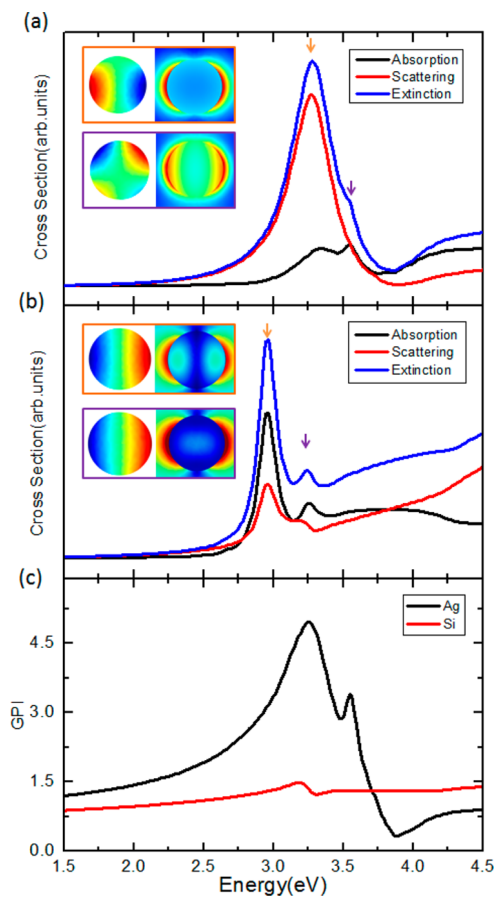

**Figure 3.** Plasmonicity in large Ag and Si NPs. (a) Calculated absorption, scattering, and extinction of an 80 nm diameter Ag nanosphere described using the Johnson and Christy permittivity.[56] The insets show the induced charge (left) and field enhancement (right) distributions for the dipolar and quadrupolar modes (top to bottom) at the frequencies indicated by arrows in the extinction spectrum. (b) Calculated absorption, scattering, and extinction of an 80 nm diameter Si nanosphere using Palik permittivity.[77] The insets show the induced charge (left) and field enhancement (right) distributions for the magnetic dipolar resonance (3.0 eV) and electric dipolar resonance (3.3 eV) (top to bottom). (c) Calculated GPI spectra for the Ag (black) and Si (red) spheres. All calculations are performed using Mie theory.

The damping clearly plays a crucial role for the GPI. This is explicitly demonstrated in the fully quantum mechanical result (eq 10), which shows that the GPI is inversely proportional to the damping. The results from the classical calculations in Figure 4 also show that damping is detrimental to plasmonicity. The GPI for the Au sphere increases strongly in a dielectric medium when the red-shifted plasmon becomes detuned from IBTs. This is also observed for the Al sphere, where the GPI increases with decreasing radiative damping for a smaller particle. The same effect is also responsible for the large GPI of the Au sphere placed in a dielectric medium compared with the Ag sphere in vacuum. The plasmon resonance for Ag has larger radiative damping because its energy is twice as large as the Au resonance and lies close to the IBTs threshold for Ag. We expect that other types of damping mechanisms, such as interfacial damping, nonlocal screening, or surface scattering will have a similar adverse effect on the GPI. IBTs are particularly prominent for transition metals, which are also known to be poor plasmonic materials. IBTs also play a detrimental role for the plasmon resonances in other noble metals such as Cu and Pt, where the GPI for 50 nm spheres is in the 1.0−1.5 range.

Figure 4c shows the calculated maximum field enhancements outside the spheres, which clearly correlate with the GPI spectra. This correlation supports our general notion that better plasmonic materials provide larger field enhancements. However, the field enhancement by itself is not an adequate metric for plasmonicity. The field enhancement around a nanostructure is typically very dependent on its shape and can be particularly large near sharp protrusions and in narrow junctions between NPs. Even for simple spheres, the field enhancement does not perfectly reflect the plasmonicity. For instance, when comparing the GPI and field enhancement for the Al spheres in Figure 4, we observe that the 50 nm Al sphere has a larger GPI but smaller field enhancement than the 25 nm sphere. A more systematic exploration of how the GPI can be used to quantify the usefulness of plasmon modes will be presented elsewhere.

**GPI for Metal and Semiconductor Clusters from TDDFT.** The numerical results presented so far concerned NPs larger than 2 nm. Here we complete the analysis by computing the GPI at the TDDFT level for both metallic and semiconductor nanoclusters in the 1−2 nm size regime. Our results indicate that the picture emerging from the jellium studies above still holds when the size of the systems is reduced, approaching the molecular limit. In particular, we consider the first two members of the series of icosahedral silver clusters $[Ag_{13}]^{5+}$ and $[Ag_{55}]^{3-}$,[59,60] where the charges have been chosen to result in a closed-shell electronic structure, as well as a tetrahedral $Ag_{20}$ cluster.[61] Then we analyze two H-passivated cubic diamond silicon nanocrystals $Si_{10}H_{16}$ and $Si_{20}H_{36}$.[62]

Both icosahedral and tetrahedral clusters are centrosymmetric and with overall sizes that overlap the smallest jellium spheres analyzed above. The study of two different geometric shapes allows us to account also for the effect of the atomic structure on the plasmonic properties.

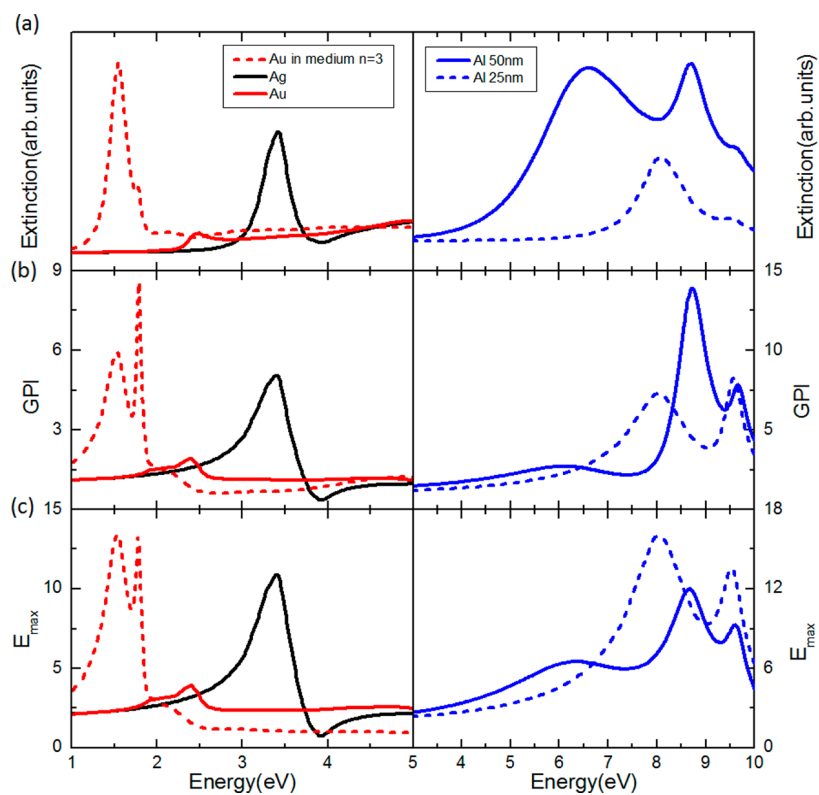

Figure 4. Plasmonicity in a large NP: Ag, Au, and Al. (a) Left: Extinction cross sections for a 50 nm diameter Ag nanosphere in vacuum (solid black) and for a 50 nm diameter Au nanosphere placed in either vacuum (solid red) or a dielectric medium of refractive index = 3 (dashed red). Right: The same for 50 nm diameter (solid blue) and 25 nm diameter (dashed blue) Al nanospheres in vacuum. (b) Corresponding GPI spectra. (c) Maximum electric-field enhancements. The calculations are performed using classical Mie theory with the Ag and Au permittivities taken from ref 56 and the Al permittivity taken from ref 78.

Figure 5a−c shows the TDDFT absorption spectra of the silver clusters. Both $Ag_{20}$ and $[Ag_{55}]^{3-}$ exhibit a dominant optical resonance around 3.1 eV. Some smaller peaks at lower energies and some intensity modulations above 3.5−4 eV are also visible. In the case of the smaller $[Ag_{13}]^{5+}$, the spectrum is more structured and the 3.1 eV resonance is replaced with a set of discrete quasi-molecular features. All absorption properties described here agree with previous theoretical studies,[59−61] although some minor differences are present due to different computational details (more in section S6 in the SI).

The TDDFT absorption spectra of the two hydrogenated silicon nanocrystals $Si_{10}H_{16}$ and $Si_{29}H_{36}$ are shown in Figure 5d and Figure 5e. They represent prototypical nonplasmonic semiconductor NPs, where excitonic effects are expected to dominate.[62]

For analysis of the GPI, we selected one peak in the spectrum of each system. More specifically, we chose the most intense peak for those systems that show a dominant (plasmon-like) peak, while we focus on the absorption edge for the remaining systems. For all the selected peaks (indicated by arrows in Figure 5), we computed the TDDFT charge density response (insets in Figure 5) and the corresponding GPI. The results are shown in Figure 6.

The GPIs identifies the metallic clusters as being more plasmonic than the silicon clusters, in agreement with the discussion above. The GPIs for the Si clusters are about an order of magnitude lower than for the Ag clusters. Given that the number of valence electrons in the sp shell is larger in the silicon clusters (four sp electrons per Si atom) than in the silver clusters (one s electron per Ag atom), we see that the GPI is more sensitive to the nature of the transition than to the number of excitable electrons. We carried out the GPI analysis also for other peaks in the low-energy region of the spectra of $[Ag_{13}]^{5+}$, $Ag_{20}$, $Si_{10}H_{16}$, and $Si_{29}H_{36}$ (see Figure S8 in the SI). These results confirm that the silicon cluster transitions have similarly low GPI values and thus no plasmonic modes. Note that the GPI values for the silver clusters are significantly smaller than those for the jellium spheres and classical particles discussed above. Within the GPI metric, this strong optical resonance is best described as incipiently plasmonic.

The variation of the GPI with increasing size of the silver clusters is nontrivial. Since the damping assumed is the same for all the systems, the difference should be related to the different Coulomb energies associated with the transitions in the various clusters. Finally, we note that the PI metric follows a similar trend as the GPI; the two metrics are compared in Figure S9 in the SI.

**Emergence of Plasmonic Behavior in a Jellium Sphere.** We now address the question of the number of electrons needed for collective plasmon modes to appear in a NP. In Figure 7a, we show the absorption spectra $\sigma_{RPA}$ and $\sigma_0$ for a gold nanosphere with diameter $D = 8$ nm as the number of electrons increase from 10 to 500. As shown in Figure 2a, quantum confinement effects are small in this case due to the relatively large NP size. The evolution of the plasmon resonances is thus determined only by the number of interacting electrons. For low electron density (<50 electrons in the particle), the absorption spectra $\sigma_{RPA}$ and $\sigma_0$ have similar resonance energies, which we refer to as $\omega_{RPA}$ and $\omega_0$ in what follows, suggesting that the excitations are essentially single-electron transitions. However, when the electron density increases (more than 100 electrons), the two absorption spectra show a considerable peak separation, signaling plasmonic behavior. The corresponding induced charge-density distributions shown in Figure 7a confirm the evolution of the absorption resonance from single-electron transitions to plasmonic excitations. At the lowest electron density (10 electrons), the induced charges are mostly confined to the interior of the NP. Conversely, for larger electron densities, the induced charge distribution becomes more surface-like and classical.

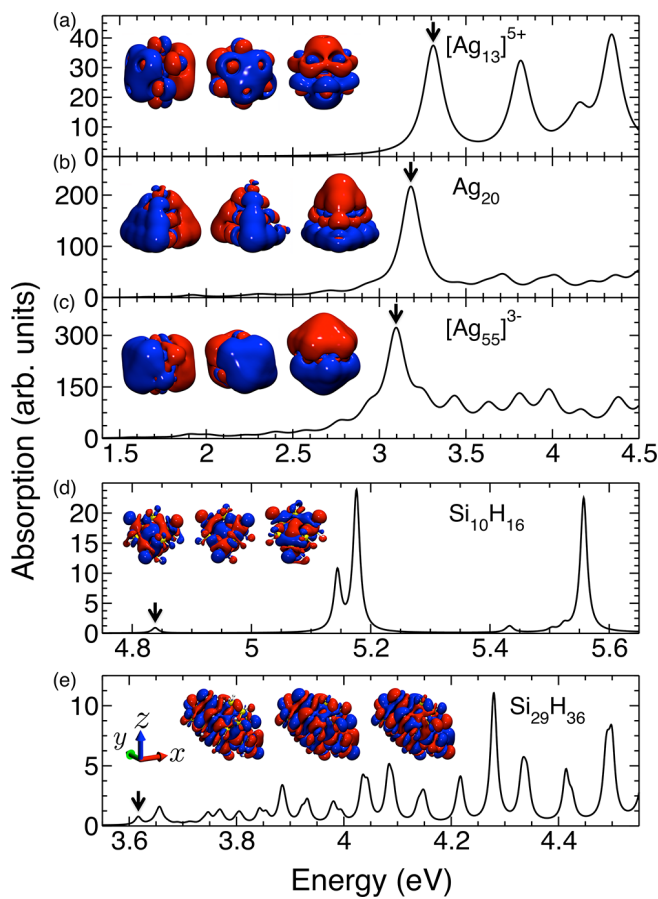

Figure 5. Absorption spectra of Ag and Si nanoclusters. The insets show the isosurface plots ($x$, $y$, and $z$ spatial polarizations, respectively) of the imaginary part of the TDDFT induced charge density response calculated for the frequencies indicated by arrows in the TDDFT absorption spectra. In particular, the excitation energies $E_{exc}$ and isosurface values iso (the same for the three polarizations) are (a) $[Ag_{13}]^{5+}$, $E_{exc}$ = 3.31 eV, iso = 0.05 bohr$^{-3}$; (b) $Ag_{20}$, $E_{exc}$ = 3.18 eV, iso = 0.05 bohr$^{-3}$; (c) $[Ag_{55}]^{3-}$, $E_{exc}$ = 3.10 eV, iso = 0.05 bohr$^{-3}$; (d) $Si_{10}H_{16}$, $E_{exc}$ = 4.84 eV, iso = 0.005 bohr$^{-3}$; (e) $Si_{29}H_{36}$, $E_{exc}$ = 3.62 eV, iso = 0.0005 bohr$^{-3}$; $x$, $y$, and $z$ coordinates relative to the induced charge density plots are shown, as reference.

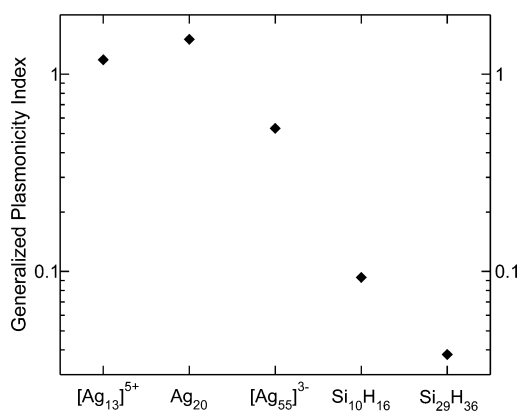

Figure 6. GPI for the Ag and Si nanoclusters considered here (log scale). GPIs are calculated from TDDFT for the peaks—one for each system—selected in Figure 5.

Calculated GPI spectra are presented in Figure 7b. As discussed above, plasmonic behavior should result in a clear peak. The GPI spectra for the particles with the smallest number of electrons (10−30 electrons) do not exhibit such a feature. A distinct peak only begins to appear for ∼50 electrons and is relatively well developed only around 100 electrons. The absence of a clear peak in the GPI spectra for low number of electrons suggests a complementary criterion for identifying plasmonic behavior based on the shape of the peak. To this end we introduce the quantity $\Delta = 1 - \frac{\eta(\omega \to 0)}{\eta_{max}}$, where $\eta(\omega \to 0)$ is the value of the GPI at zero frequency, and $\eta_{max} = \eta(\omega_{GPI})$, where $\omega_{GPI}$ is the energy of the GPI resonance (which has coincided with the RPA plasmon resonance $\omega_{RPA}$ in all cases considered so far). In the zero-frequency limit, no plasmons are excited, and the induced fields screen the external fields almost perfectly, leading to $\eta(\omega \to 0) = 1$ except for very small systems with large quantum size effects (Figure S2). A value of $\Delta$ close to 1 implies a well-defined GPI resonance and is thus a signature of plasmonic behavior. In Figure 7c, we plot $\omega_0$ and $\omega_{RPA}$ (top), $\eta_{max}$ (middle), and $\Delta$ (bottom) for the different nanospheres. The evolution of $\omega_{RPA}$ clearly shows a square root trend as a function of electron density, as predicted by classical theory, while the peaks of $\omega_0$ do not change significantly. The middle panel in Figure 7c shows a monotonic increase of $\eta_{max}$ with electron density, reaching $\eta_{max} > 2$ when the electron number exceeds 100. In the bottom panel of Figure 7c, we plot the $\Delta$ values obtained from the GPI spectra (Figure 7b) and find that $\Delta$ also increases monotonically with electron density. The results in Figure 7 allow us to formulate two equivalent criteria for plasmonic behavior: $\eta_{max} > 2$ and $\Delta > 0.5$. All metrics introduced in Figure 7c are consistent with plasmonic behavior emerging for NPs with more than 100 electrons in the present $D$ = 8 nm NP. More details on how the $\Delta$ quantity is related to the separation between $\omega_{RPA}$ and $\omega_0$ and to the $\lambda$ dispersion of the $\omega_\lambda$ resonance in the scaling procedure as well as its size dependence are presented in Section 2 of the SI. Overall, the GPI and the corresponding $\Delta$ quantity provide an intuitive approach for quantifying plasmonic behavior in a nanostructure.

**Plasmons in Polycyclic Aromatic Hydrocarbons and Nanostructured Graphene.** Graphene nanostructures exhibit low-energy plasmons that strongly depend on the level of electrical doping and geometry.[63] When the size of a structure is reduced so that it contains only a few carbon atoms, one encounters polycyclic aromatic hydrocarbons (PAHs), which were predicted[38] and experimentally demonstrated[37] to exhibit qualitatively similar behavior as graphene. These discoveries have stimulated a variety of studies and created a subfield of quantum plasmonics research, molecular plasmonics. Classification of the optical modes supported by PAHs initially relied upon the $\chi$ versus $\chi^0$ criterion discussed above,[38,64] but as demonstrated below, the GPI provides a more universal criterion to identify plasmonic behavior in these systems as well.

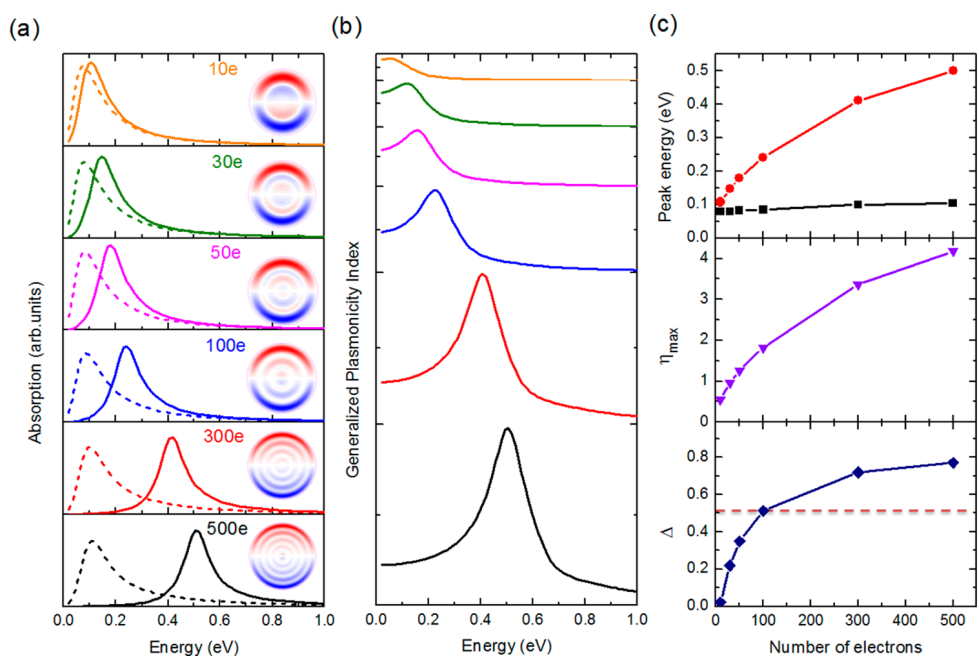

Figure 7. Plasmonicity in a metallic NP: dependence on electron density. (a) Evolution of absorption spectra, $\sigma_{RPA}$ (solid) and $\sigma_0$ (dashed), for increasing number of electrons. The insets show the charge-density distributions associated with the plasmons. (b) GPI spectra as a function of photon energy for the nanospheres in (a). (c) (Top) Evolution of resonance energies (peaks) of $\omega_{RPA}$ (red) and $\omega_0$ (black). (Middle) Evolution of the GPI values at the resonance $\eta_{max}$. (Bottom) Evolution of $\Delta$; see text for definition. The dashed horizontal line is $\Delta = 0.5$. The diameter of the jellium sphere is $D = 8$ nm in all cases.

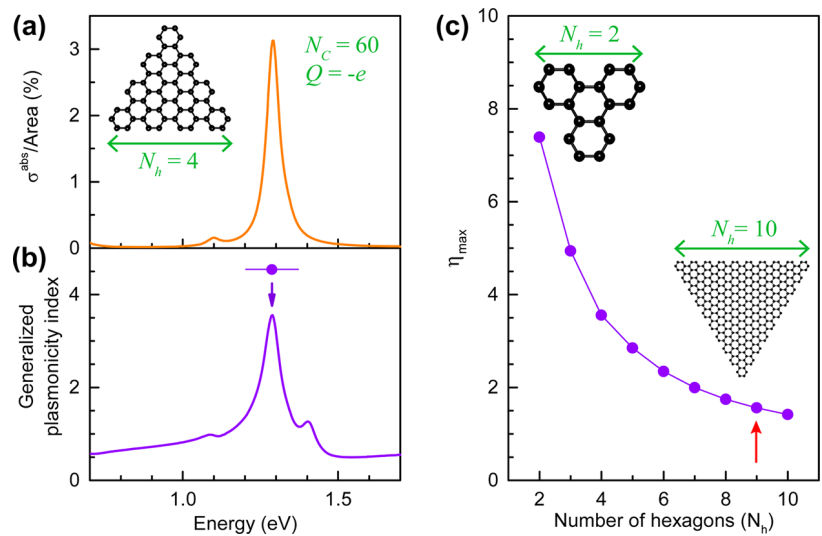

Figure 8. (a) Absorption spectrum for the singly charged $N_h = 4$ hexagon (containing $N_C = 60$ carbon atoms) normalized to the nanotriangle area. (b) The corresponding GPI spectrum. (c) Size evolution of the GPI at the plasmon resonances of armchair graphene nanotriangles (see carbon atomic structures in the insets) doped with a single excess electron. The size $N_h$ denotes the number of benzene hexagonal rings spanning each side. The assumed damping is 25 meV. Edge carbons are passivated with hydrogen atoms (not shown).

In Figure 8, we present a GPI analysis of the dominant low-energy optical features in triangular-shaped PAHs of increasing size (measured by the number of benzene hexagonal rings $N_h$ along the triangle side), doped only with one excess electron. We calculate the optical response of these systems following a previously reported RPA approach (tight-binding RPA, TB-RPA),[65,66] using valence electron wave functions from a tight-binding model with one spin-degenerate $p_z$ orbital per carbon site. A characteristic TB-RPA absorption spectrum (Figure 8a for a PAH of $N_h = 4$ consisting of $N_C = 60$ carbon atoms) clearly reveals a distinct optical resonance around 1.3 eV. The corresponding GPI spectrum (Figure 8b) also displays a maximum at the energy of this peak. The dependence of the GPI maximum $\eta_{max}$ on PAH size (Figure 8c) reveals a clear evolution from plasmon-like behavior (GPI > 7) for the smallest molecule under consideration (triphenylene, with $N_h = 2$ and only $N_C = 18$ carbon atoms) to less plasmonic character for large sizes as expected.[37,38] The molecular plasmon in a PAH is enabled by a change of the electronic structure associated with addition (or removal) of electrons, while for graphene, it is the injected electrons or holes that make up the electron gas sustaining the plasmon. As the structure becomes larger, the electron density becomes smaller, resulting in less plasmonic behavior, as also shown for the jellium spheres in Figure 7. We thus conclude that very small PAHs can display plasmons, even when they are singly charged. In this respect, we note that the effective number of electrons contributing to the plasmonic strength is augmented in graphene by the nonparabolic band structure of its conduction electrons, so that a comparatively small number of doping charge carriers produces a comparatively larger response than in metals (i.e., systems with nearly parabolic dispersion). This effect is quantified by the fact that the effective number of charge carriers contributing to the response is roughly given by the geometrical average of the number of doping charges times the number of carbon atoms, as it has been previously investigated for small PAHs.[63] This explains why plasmons are sustained in these systems even if they have a small number of electrons compared with the jellium spheres considered in Figure 7.

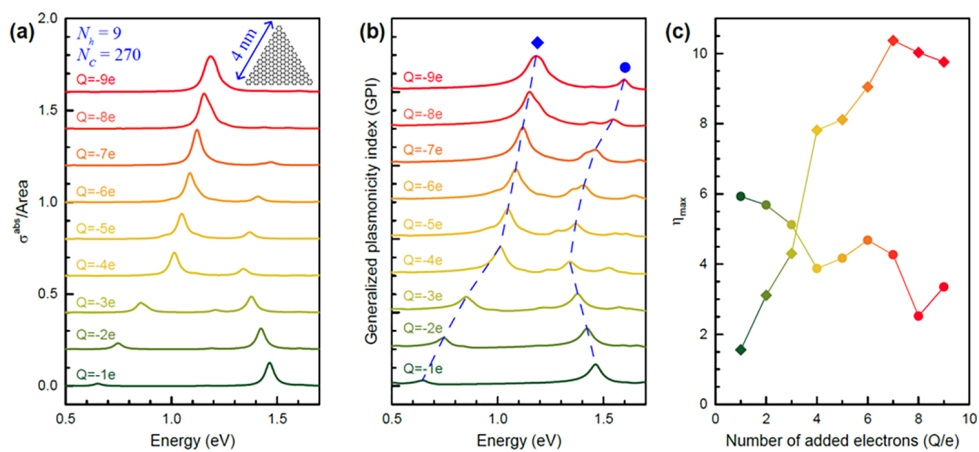

Figure 9. (a) Absorption spectra of a 4 nm armchair graphene nanotriangle (containing $N_C$ = 270 carbon atoms and $N_h$ = 9 benzene rings along a side, see inset), shown as the number of added electrons increases from 1 to 9. (b) Evolution of the GPI associated with the spectra shown in (a). The dashed lines are guides to the eye, revealing the evolution of two prominent modes supported by the structure as the doping charge is varied. (c) GPI maxima for the two resonances highlighted in (b), where the square (circular) symbols correspond to the lower-energy (higher-energy) mode.

In Figure 9, we consider a graphene nanotriangle of fixed size ($N_h$ = 9, with $N_C$ = 270 carbon atoms). The TB-RPA absorption spectra shown in Figure 9a display an interesting evolution as the number of doping electrons increases. The spectra exhibit two distinct resonances (low- and high-energy features marked with dashed curves as guides to the eye in Figure 9b). The low-energy feature discussed in Figure 8 (i.e., the leftmost feature of the $Q/e = -1$ spectrum in Figure 9a, corresponding to the red arrow in Figure 8c) blue shifts when adding electrons, following a similar frequency dependence $\sim Q^{1/4}$ on doping charge $Q$ as in extended graphene.[65] As inferred from the analysis of Figure 8, the structure is sufficiently large to prevent plasmonic behavior with only one doping electron. However, as revealed by the GPI spectral analysis (Figure 9b), the low-energy feature becomes more plasmonic with increasing electron density. In contrast, the high-energy feature in the plotted spectra does not exhibit such behavior and becomes less plasmonic with increasing electron density. This feature is associated with the HOMO−LUMO IBT found in the electronic structure of armchair-edged graphene nano-triangles[32,55] and is quenched as the LUMO states are populated with additional electrons due to Pauli blocking[67] (see Section S7 in the SI). Consequently, the GPI peak of this high-energy feature mostly decreases with doping, unlike the low-energy feature, which increases until it saturates at GPI ∼ 10.

## CONCLUSIONS

We have presented a unified theoretical approach for the identification of collective plasmon excitations that can be applied to different complementary descriptions ranging from *ab initio* methods to classical electrodynamics of continuous media. In particular we introduced the generalized plasmonicity index (GPI) as a metric for the evaluation of plasmonic behavior. The GPI is directly expressed in terms of the light-induced electron density change, a quantity that is readily available in virtually any computational scheme. Using this metric, we studied the emergence of plasmonic behavior with increasing numbers of electrons in a spherical jellium particle. Specifically, we demonstrated that for a quantum plasmon mode to display classical behavior for a NP of nanometer size ($D$ = 8 nm), around 100 electrons are needed. We also showed that the GPI can be used in conjunction with classical electromagnetic simulations to discriminate between plasmonic and photonic modes and that it represents a metric for classifying the quality of a classical plasmon.

Moving toward the molecular scale, we demonstrated that the GPI also provides a physically sound classification of the plasmonic character of optical excitations in noble-metal clusters and polycyclic aromatic hydrocarbons. The presented findings introduce a systematic classification of excitations in terms of their plasmonic character that can disclose currently *incognito* plasmonic-based physics in optically active nanomaterials.

## METHODS

**Jellium Model.** The jellium spheres were modeled with an electron density and background permittivity of 9.1 as appropriate for Au. The optical spectra were calculated using the time-dependent local density approximation (TDLDA), which amounts to first calculating the electronic structure using LDA and then solving for the optical response using the full RPA with a suitably chosen functional to include interactions between electrons.[26,46,47] The electronic ground states were calculated from the Kohn–Sham equations:

$$\left[-\frac{\hbar^2}{2m}\nabla_r^2 + V_{eff}(r)\right]u_{lk}(r) = E_{lk}u_{lk}(r) \quad (M1)$$

where $m$ is the electron mass, $u_{lk}(r)$ the radial wave functions, and $E_{lk}$ the corresponding eigenenergies for angular and principal quantum wave numbers $l$ and $k$. $V_{eff}(r)$ is the effective one-electron potential at position $r$: $V_{eff}(r) = V_{ext}(r) + V_{xc}[n(r)] + V_h[n(r)] + \frac{\hbar^2 l(l+1)}{2mr^2}$, where $V_{ext}$ is a constant external potential adjusted so that the work function of gold is reproduced,[68] $V_{xc}$ is the exchange–correlation potential,[68] $n(r)$ is the local electron density, and $V_h$ is the Hartree potential.

**Random-Phase Approximation.** After obtaining the eigenstates and eigenvalues, the absorption spectra were computed using the RPA.[26,69] The induced charge distribution can be calculated from

$$\delta n(\mathbf{r}, \omega) = \int \chi(\mathbf{r}, \mathbf{r}', \omega) v_{ext}(\mathbf{r}', \omega) d\mathbf{r}' = \int \chi^0(\mathbf{r}, \mathbf{r}', \omega) v_{tot}(\mathbf{r}', \omega) d\mathbf{r}' \quad (M2)$$

where $\chi(\mathbf{r},\mathbf{r}',\omega)$ is the density response function kernel and $\chi_0(\mathbf{r},\mathbf{r}',\omega)$ is the irreducible response function for independent electrons. Also, $v_{tot}(\mathbf{r},\omega) = v_{ext}(\mathbf{r},\omega) + \lambda v_{ind}(\mathbf{r},\omega)$ is the total potential consisting of the external potential $v_{ext}(\mathbf{r},\omega)$ and the induced potential $v_{ind}(\mathbf{r},\omega)$, which is defined as $v_{ind}(\mathbf{r},\omega) = \int d\mathbf{r}' K(\mathbf{r},\mathbf{r}') \delta n(\mathbf{r}',\omega)$ with $K(\mathbf{r},\mathbf{r}')$ being the Coulomb interaction kernel. As mentioned earlier, we introduced a scaling parameter $\lambda$, denoting the fraction of induced Coulomb potential $v_{ind}(\mathbf{r},\omega)$ that enters into the total potential $v_{tot}(\mathbf{r},\omega)$. When $\lambda = 0$, the induced potential is turned off and $\chi(\mathbf{r},\mathbf{r}',\omega) = \chi^0(\mathbf{r},\mathbf{r}',\omega)$. In this case, the Coulomb interaction between globally distributed induced charges is ignored, and only single-electron transitions are considered, as illustrated in the right-hand side of Figure 1. When $\lambda = 1$, we recover the full RPA framework. From the induced charge density, it is convenient to introduce the local dipolar polarizability defined as $\alpha(r, \omega) \equiv r^2 \cdot \delta n(r,\omega)$. The RPA equation can then be expressed in terms of the dipolar polarizability as

$$\alpha(r, \omega) = \alpha_0(r, \omega) + \lambda \int dr' P(r, \omega) \alpha(r', \omega) \quad (M3)$$

in which $\alpha_0(r, \omega) = \int \chi^0_{l=1}(r,r', \omega) v_{ext}(r', \omega) dr'$ and $P(r, \omega)$ is proportional to the dipolar component of the independent response function for the independent charge susceptibility $\chi^0_{l=1}$ in the quasistatic limit.[26] By introducing the dipolar polarizability $\alpha(\omega) = \frac{4\pi}{3} \int dr' r' \alpha(r', \omega)$ the dipolar absorption is obtained through the relation:

$$\sigma_{abs}(\omega) = \frac{4\pi\omega}{c}\text{Im}[\alpha(\omega + i0^+)] \quad (M4)$$

$$\sigma_0(\omega) = \frac{4\pi\omega}{c}\text{Im}[\alpha_0(\omega + i0^+)] \quad (M5)$$

where $c$ is the speed of light. Eq M4 can be applied to all systems where the RPA framework holds, and thus, it is not limited to the jellium model studied here.

**Tight-Binding RPA Description of PAHs.** We followed a previously described method[65,66] to obtain the single-particle, valence-electron wave functions of graphene-related structures using a tight-binding model that incorporates a single out-of-plane p orbital per carbon site $l$, with a nearest-neighbor hopping energy of 2.8 eV (edge atoms are considered to be hydrogen-passivated and undergo the same hopping with their nearest-neighbors). We then adapted the RPA formalism to express the response in this discrete basis set, which results in an induced charge density $\rho_l^{ind}$ at each site $\mathbf{R}_l$, as well as induced and external potentials $\phi_l^{ind}$ and $\phi_l^{ext}$. This allows us to calculate the absorption cross-section (neglecting retardation) for a unit external electric field as $\sigma_{abs}(\omega) = \Sigma_l(4\pi\omega/c)\mathbf{R}_l\text{Im}\{\rho_l^{ind}\}$, and the GPI as

$$\eta = \frac{|\Sigma_l \rho_l^{ind}(\phi_l^{ind})^*|}{|\Sigma_l \rho_l^{ind}(\phi_l^{ext})^*|} \quad (M6)$$

Incidentally, this TB-RPA approach yields results in excellent agreement with full-electron TDDFT down to small structures,[38] such as the triphenylene molecule ($N_C = 18$ carbon atoms) considered above.

**TDDFT-Based Calculations.** The electronic structures and absorption spectra of metallic and semiconductor nanoclusters were simulated using the Quantum ESPRESSO (QE) suite of codes,[70] based on TDDFT within periodic boundary conditions. The PBE[71] generalized gradient approximation (GGA) to the exchange–correlation (xc) functional was adopted. No Brillouin zone sampling was performed, given these are finite systems. Vanderbilt ab initio ultrasoft (US) scalar-relativistic pseudopotentials (PPs)[72] were used to describe the core electrons and nuclei; however, the 4d electrons of Ag were explicitly included. The single-particle wave functions were expanded in plane waves up to a kinetic energy cutoff of 45 Ry for Ag clusters and 28 Ry for Si nanocrystals. Consistent with these values and the use of US-PPs, the kinetic energy cutoff for the charge densities was 540 and 280 Ry, respectively. The calculated GPI was not much sensitive to the xc functional used: test calculations performed on $Ag_{20}$, using LDA and hybrid B3LYP xc functionals, as well as with norm-conserving PPs on $Ag_{20}$ and $[Ag_{13}]^{5+}$, were very similar to the results from PBE-US-PPs. All simulations exploit periodically repeated supercells, each containing the system and a suitable amount of vacuum (12 Å at least) in the three spatial directions, in order to separate adjacent replica and to avoid spurious interactions. A compensating jellium background was inserted to remove divergences in the charged systems. Since this may cause errors in the potential if the cell is not large enough, we run a test by calculating GPIs for $[Ag_{13}]^{5+}$ with a larger cell (amount of vacuum increased by 50%). The picture provided by the resulting GPI was the same. The ideal icosahedral silver clusters were created by adding Mackay shells according to the homonymous protocol,[73,74] using the experimental bulk value for the bond length. Silicon nanocrystals were obtained from the bulk silicon whose dangling bonds were passivated with hydrogen atoms. All the atomic structures were relaxed by using QE with the PBE-GGA xc functional. The TurboTDDFT code,[75] also distributed within QE package, was employed to compute the optical absorption spectra, as detailed elsewhere,[39] and the response charge densities of Ag and Si NPs. This code implements the Liouville–Lanczos approach to linearized TDDFT[76] in the frequency domain; it is optimized to treat relatively large systems and enables calculation of spectra over a wide energy range. For each polarization of the external electric field, 10,000 Lanczos iterations were performed and then averaged over the three spatial coordinates to obtain the spectra.

## ASSOCIATED CONTENT

**Supporting Information**

The Supporting Information is available free of charge on the ACS Publications website at DOI: 10.1021/acsnano.7b03421.

> Plasmonicity index discussion (S1); quantitative methods for distinguishing a plasmon from a single particle transition (S2); decomposition of the RPA excitation energy (S3); the GPI is proportional to $E_{plas}$ at resonance (S4); the GPI in the quasistatic limit (S5); more on ab initio absorption spectra of clusters, induced charge densities, PI and GPI (S6); and further details into the plasmonic nature of PAHs excitations (S7) (PDF)

## AUTHOR INFORMATION

**Corresponding Authors**
*E-mail: nordland@rice.edu.
*E-mail: yiaimer@gmail.com.
*E-mail: javier.garciadeabajo@icfo.es.

*E-mail: stefano.corni@unipd.it.
ORCID
Alejandro Manjavacas: 0000-0002-2379-1242
Stefano Corni: 0000-0001-6707-108X
Emily A. Carter: 0000-0001-7330-7554
Peter Nordlander: 0000-0002-1633-2937

**Present Address**
◇Laboratory for Nanophotonics and the Department of Physics and Astronomy, MS61, Rice University, Houston TX 77005, United States

**Author Contributions**
P.N. and J.G.A. conceived the project. R.Z. and H.Z. performed the jellium calculations and developed the GPI metric. A.C., S.C., and E.M. designed the cluster calculations. S.C. and L.B. derived eq 10. L.B. and A.C. performed the Ag cluster calculations. J.DC. performed the TB-RPA calculations.

**Notes**
The authors declare no competing financial interest.

## ACKNOWLEDGMENTS
This work was supported in part by the Robert A. Welch Foundation under grant C-1222 (P.N.), the Air Force Office of Scientific Research under grant MURI FA9550-15-1-0022 (P.N. and E.A.C.), the Spanish MINECO (MAT2014-59096-P, Fundació Privada Cellex, and SEV2015-0522), the European Commission (Graphene Flagship CNECT-ICT-604391 and FP7-ICT-2013-613024- GRASP), the European Union through the ERC Consolidator Grant 681285 TAME-Plasmons (S.C.), and the European Center of Excellence 'MaX-Materials at the exascale', grant 676598 (E.M.). A.M. acknowledges financial support from the Department of Physics and Astronomy and the College of Arts and Sciences of the University of New Mexico. C.M.K. acknowledges support by a fellowship within the Postdoc Program of the German Academic Exchange Service (DAAD).